\documentclass[pre,showpacs,12pt]{revtex4}

\usepackage[english]{babel}
\usepackage{graphicx}

\begin{document}

\begin{titlepage}

\title{ Thermodynamics of Electrolytes on Anisotropic Lattices}

\author{Vladimir Kobelev \footnote{Current address: Dept. of Materials
Science and Engineering, UIUC, Urbana, IL 61820}, Anatoly B.
Kolomeisky}

\affiliation{Department of Chemistry, Rice University, Houston,
Texas 77005}

\author{Athanassios Z. Panagiotopoulos}

\affiliation{Department of Chemical Engineering, Princeton
University, Princeton, NJ 08544}


\begin{abstract}

The phase behavior of ionic fluids on simple cubic and tetragonal
(anisotropic) lattices has been studied by grand canonical Monte
Carlo simulations.  Systems with both the true lattice Coulombic
potential and continuous-space $1/r$ electrostatic interactions
have been investigated.  At all degrees of anisotropy, only
coexistence between a disordered low-density phase and an ordered
high-density phase with the structure similar to ionic crystal was
found, in contrast to recent theoretical predictions. Tricritical
parameters were determined to be monotonously increasing functions
of anisotropy parameters which is consistent with theoretical
calculations based on the Debye-H\"uckel approach. At large
anisotropies a two-dimensional-like behavior is observed, from
which we estimated the dimensionless tricritical temperature and
density for the two-dimensional square lattice electrolyte to be
$T^*_{tri}=0.14$ and $\rho^*_{tri} = 0.70$.

\end{abstract}

\maketitle
\end{titlepage}

\section{Introduction}

After controversial  experimental results on the nature of ionic
criticality \cite{pitzer,japas,weingartner01} the studies of
critical phenomena in ionic fluids received a fresh impetus, and
over the last decade substantial progress has been made
\cite{weingartner01,levin02}. However, a complete thermodynamic
description of electrolyte systems is still unavailable.

The simplest and  most frequently employed model system for both
theoretical and Monte Carlo investigations of Coulombic systems is
the restricted primitive model (RPM), which is a gas of equisized
hard-sphere ions carrying positive and negative charges of equal
magnitude. From the theoretical point of view, a reasonable
description of the RPM in the critical region has been obtained at
a mean-field level using integral equations approach
\cite{stell95,yeh96,blum02} and Debye-H\"{u}ckel theory
\cite{levin02,levin96}. While theoretical predictions of the
critical parameters  \cite{levin96,blum02,kulinskii} agree
reasonably well with those obtained in computer simulations
\cite{panagiotopoulos99,caillol02}, recent Monte Carlo
investigations of ionic systems indicate a non-classical
Ising-like asymptotic critical behavior \cite{lujiten02}.
Moreover, predictions of different mean-field theories for charge-
and size-asymmetric  primitive  systems  differ significantly from
each other \cite{zuckerman01,banerjee02,artyomov03,netz,sabir}. In
the absence of real experimental data, numerical simulations
provide an important criterion for testing different theoretical
methods \cite{camp99,panagiotopoulos02,yan02,cheong03}.

In recent years, a new theoretical approach to study the
thermodynamics and criticality of Coulomb systems based on the
investigation of corresponding lattice models, has been introduced
\cite{dickman99,kobelev02-1,brognara,ciach01,kobelev02-2,artyomov03}.
Being obviously less realistic than their continuous-space
counterparts, lattice models are much easier to handle
analytically, and the information they provide is crucial for
understanding the physics of criticality. Moreover, finely
discretized lattices with lattice spacing several times smaller
than the ion size present substantial computational advantages
over continuous space simulations, while producing almost
identical critical parameters and phase diagrams
\cite{panagiotopoulos99}.  By contrast, when the ionic diameter is
equal to the spacing of the underlying lattice, the phase diagram
is drastically different.  The symmetry of simple cubic or
body-centered cubic lattices allows for unequal charge
distribution between two identical sublattices, yielding an
ordered structure similar to an ionic crystal. The competition
between electrostatic interactions of the two charged sublattices
and the entropy of charge separation leads then to order-disorder
phase transitions and a tricritical point \cite{dickman99,
panagiotopoulos99,kobelev02-1,brognara}.

A systematic study  of the lattice restricted primitive model
(LRPM) based on the Debye-H\"uckel approach has shown that, in
accordance with Monte Carlo simulations, a Coulombic-governed
gas-liquid phase transitions seen in continuum electrolytes is
totally suppressed on simple cubic  and body-centered cubic
lattices due to the formation of the thermodynamically more stable
charge-ordered phase \cite{kobelev02-1}. Ciach and Stell
\cite{ciach01} argued that, unlike non-ionic fluids, in charged
systems the most important are short-range fluctuations.
Consequently, they studied a LRPM supplemented with additional
short-range interactions between ions and concluded that at some
strength of the short-range attraction both the normal critical
and tricritical point can become thermodynamically stable
\cite{ciach01}. This leads to  complex phase diagrams with two
distinct phase transitions, one between  gas and  disordered
liquid phase, and another one  between disordered liquid and a
charge ordered phase. Recent Monte Carlo studies have confirmed
qualitatively most of these theoretical predictions but further
investigations are required \cite{diehl03}.

At low dimensions  Coulombic interactions are stronger than in
three-dimensional systems, and the possibility for gas-liquid
phase transitions increases. This idea has been used recently in
the  investigation of anisotropic lattice models of electrolytes
using Debye-H\"{u}ckel method \cite{kobelev02-2}. In this study
the anisotropy mimicked the lowering of the spatial
dimensionality. It has been found, that for strongly anisotropic
lattices gas-liquid phase coexistence is restored. However, this
theory neglects an ion clustering phenomena, which are important
in  charged particles  systems \cite{levin96,levin02}. Thus these
theoretical predictions should be tested by computer  simulations.

Current  Monte Carlo  studies of lattice RPM in $d=3$ dimensions
employ the continuous space $1/r$ potential
\cite{panagiotopoulos99,dickman99,diehl03}. However, because of
the discrete lattice symmetry, the correct lattice Coulomb
potential differs from $1/r$ at short distances and approaches it
asymptotically at large distances.  This raises the question of
how using the correct electrostatic interactions will affect the
thermodynamics  and critical behavior of ionic systems. Indeed,
the lattice correction to $1/r$ potential decays rapidly with
distance and thus it can be viewed in a sense as an additional
short-range interaction. In order to understand correctly  the
underlying mechanisms of phase transitions in  Coulombic systems,
lattice models should be investigated with the corresponding
lattice potentials.

In this paper we study the RPM  on tetragonal lattices using grand
canonical Monte Carlo simulations and histogram reweighting
technique \cite{ferrenberg}. The correct lattice Coulomb potential
is used and an Ewald-like summation is utilized to account for the
long-range nature of electrostatic interactions. The paper is
organized in the following way. In Sec. II the details of the
model and simulation method are given. The results and discussions
are presented in Sec. III, and summary and conclusions are drawn
in Sec. IV. Analytical expressions and the details of lattice
potential calculations are given in Appendix.

\section{Anisotropic Lattice Model and Simulations Method}

The system we have studied consists of $2N$ ions,  half of them
carrying charge $q$ and half $-q$, positioned on a simple
tetragonal lattice with lattice parameters ratio (degree of
anisotropy) $\alpha = b/a$. The pairwise interaction potential of
two ions separated by distance $r_{ij}$ has the form (for details
see Appendix A)
\begin{equation} \label{fiofr}
    U_{ij}=\frac{1} {4\pi^2 Db}\int_{-\pi}^{\pi} \frac{e^{i \mathbf{k r}} d^3 \mathbf{k}}{2+1/\alpha^2 - (\cos k_1 + \cos k_2 + 1/\alpha^2 \cos k_3)},
\end{equation}
where $D$ is the dielectric constant of the structureless solvent.
Reduced quantities are obtained by scaling with the energy of the
strongest ion-ion interaction in two neighboring $xy$ planes in
the continuum limit,  $E_0  = q^2/Db$, and the unit cell volume,
$v_0 = a^2b$, namely,
\begin{equation}
T^* = \frac{k_B T}{E_0}, \;\;\; \rho^* = \frac{2N a^2b}{V},
\end{equation}
where $V$ is the volume of the system.

Since the Coulomb potential is  long ranged, one needs to account
for interactions with particles in all images of the periodic box
used in the simulations. In the case of continuous-space $1/r$
interactions the standard method to achieve this is the Ewald
summation (see e.g. Ref.\cite{frenkel, deLeeuw}). However, no
similar method is available for the lattice Coulomb potential.
Therefore, we propose the following approximate scheme, which is
based on the asymptotic properties of the anisotropic lattice
potential (\ref{fiofr}).

It can be shown that  the potential  (\ref{fiofr}) for finite
$\alpha$ at large distances behaves  as
\begin{equation} \label{asympt}
U_{ij}({\bf r}=(x,y,z)) \sim \frac{\alpha}{\sqrt{x^2 + y^2+ \alpha z^2}}
\end{equation}
For this potential, an Ewald-like sum  can be constructed along
the lines of the derivation in Ref.\cite{deLeeuw}, which yields
the ``continuum'' part of the energy of interactions of periodic
system of $2N$ particles
\begin{eqnarray} \label{ewald} \nonumber
    E_1 =\sum_{all\, space}\frac{1}{|{\bf r}^\prime_{ij}|} &=&  \sum_{i,j=1}^{2N} q_i q_j \left(\sum_{{\bf n}^\prime}\frac{\alpha^2 \exp(-\pi^2|{\bf n}^\prime|^2/\eta^2 + 2\pi i {\bf n}^\prime {\bf r}^\prime_{ij})}{2 \pi |{\bf n}^\prime|^2} + \frac{1}{2} \sum_{{\bf n}^\prime} \frac{ {\rm erfc}(\eta |{\bf r}_{ij} + {\bf n}^\prime|)}{|{\bf r}^\prime_{ij} + {\bf n}^\prime|} \right) \\
 &+& \sum_{ |{\bf n}^\prime| \ne 0}\left(\frac{ {\rm erfc}(\eta |{\bf n}^\prime|)}{|{\bf n}^\prime|} + \alpha^2 \frac{\exp(-\pi^2|{\bf n}^\prime|^2/\eta^2)}{\pi^2 |{\bf n}^\prime|^2} \right) - \frac{\eta}{\sqrt{\pi}}\sum_{i=1}^{2N} q_i,
\end{eqnarray}
where $r^\prime = (x/\alpha,y/\alpha, z)$, $n^\prime = (\alpha n_x, \alpha n_y,n_z)$, $0 \le n_x,n_y,n_z \le 2\pi$, and ${\rm erfc(x)}$ is the complimentary error function. Here  conducting boundary conditions have been utilized, which are less sensitive to finite-size effects \cite{panagiotopoulos02}.   As usual, the real-space damping parameter $\eta$ must be chosen in a such way that all sums in (\ref{ewald}) converge fast. In our simulations we used $\eta=5$, and the Fourier-space sums were restricted to 518 wave vectors.

The difference between the  correct lattice potential
(\ref{fiofr}) and its asymptotic limit (\ref{asympt}) decays
rapidly with distance: see Table 1. Therefore, to  obtain the
total energy of electrostatic interactions, the lattice correction
is added for ions in elliptic shells up to certain distance,
\begin{equation}
    E_{total} = E_1 +\sum_{|{\bf r}^\prime/L|< n_{max}}\left(U_{ij}(r) - \frac{1}{|{\bf r}^\prime_{ij}|}\right).
\end{equation}

\begin{table}[t] \caption{Dimensionless lattice Coulomb potential $U_{lattice}$ (\ref{fiofr}) and its asymptotic limit $U_{cont}$ given by Eq.(\ref{asympt})} \label{table.3dlat}
  \begin{center}
  \begin{tabular}{|c|c|c|c|c|c|c|} \hline
     crystallographic   & \multicolumn{2}{|c|}{$\alpha$=1}   &  \multicolumn{2}{|c|}{$\alpha$=4}   & \multicolumn{2}{|c|}{$\alpha$=10} \\
      \cline{2-7} indexes& $U_{lattice}$ & $U_{cont}$ &$U_{lattice}$ & $U_{cont}$ & $U_{lattice}$ & $U_{cont}$ \\
 \hline
     \, [1 0 0]      & 1.0815164 & 1.0000000 & 3.192933  & 4.000000 & 4.949445 & 10.00000       \\
     \, [0 0 1]      & 1.0815164 & 1.0000000 & 1.028176  & 1.000000 & 1.006922 & 1.000000       \\
    \,  [10 0 0]     & 0.1002578 & 0.1000000 & 0.410816  & 0.400000 & 1.066964 & 1.000000       \\
    \,  [0 0 10]     & 0.1002578 & 0.1000000 & 1.000167  & 0.100000 & 0.100015 & 0.100000       \\
     \, [10 10 10]   & 0.0577029 & 0.0577350 & 0.094250  & 0.094281 & 0.099021 & 0.099015       \\
     \, [100 0 0]    & 0.0100003 & 0.0100000 & 0.040009  & 0.040000 & 0.100129 & 0.100000       \\
     \, [0 0 100]    & 0.0100003 & 0.0100000 & 0.010003  & 0.010000 & 0.010041 & 0.010000       \\
     \, [100 100 100]& 0.0057736 & 0.0057733 & 0.009431  & 0.009428 & 0.009924 & 0.009901   \\ \hline
   \end{tabular}
   \end{center}
\end{table}

We performed grand canonical Monte Carlo simulations on cubic
boxes of  length $L$ stretched in $z$ direction by a factor of
$\alpha$, under periodic boundary conditions. Distance-biased
algorithm was employed for insertions and removals of pairs of
unlike ions at each time step in order to facilitate acceptance,
following Ref.\cite{orkoulas94}. To analyze the simulation data
and obtain the coexistence curves, multi-histogram reweighting
techniques \cite{ferrenberg} have been used.

\section{Results and Discussion}

To reduce possible finite-size effects,  we used a system size
$L^* \equiv L_{x,y}/a = L_z/b = 12$ for the isotropic lattice
($\alpha=1$) and $\alpha=1.2$, and $L^*=16$ for stronger
anisotropies. Short initial runs were used to get a rough estimate
of the location of the tricritical point. After that, typical runs
involved $1-3 \times 10^7$ Monte Carlo steps for equilibration and
$2-7 \times 10^8$ for production. We did not try to locate the
second order transitions N\'eel line and restricted ourselves with
obtaining approximate location of the tricritical point by simple
linear extrapolation of the coexistence lines, as expected for
$d=3$ tricriticality.

In order to separate the effect of using  the correct lattice
potential in simulations, we studied first the phase coexistence
in LRPM on a simple cubic lattice: see Fig.1. On a simple cubic
lattice, Coulomb gas with $1/r$ potential is known to phase
separate, with two coexisting phases being a low-density
disordered phase and an ordered high-density ionic crystal-like
phase \cite{panagiotopoulos99, dickman99}. The tricritical
parameters  estimates for the  system size $L^* = 12$, obtained by
Panagiotopoulos and Kumar \cite{panagiotopoulos99}, are $T^*_{tri}
= 0.15\pm 0.01,\,\rho^*_{tri} = 0.48\pm 0.02$
\cite{panagiotopoulos99}.  For the same system size, taking into
account the lattice correction to the potential yields  $T^*_{tri}
= 0.22,\,\rho^*_{tri} = 0.48$ with the same accuracy. Thus  the
tricritical density remains the same within the error limits, but
the temperature increases by the factor of $1.5$. Qualitatively,
the increase in the tricritical temperature should be expected
since lattice Coulomb interactions are stronger than $1/r$ at
short distances, which leads to a higher stability of dense
phases. Nevertheless, the tricritical point is still significantly
lower (by 50\%) than the predictions of the Debye-H\"uckel theory
of lattice electrolytes $T_{tri}\approx 0.365$ \cite{kobelev02-1},
but is about 20\% higher than the results $T_{tri}\approx 0.202$
from the hierarchical reference theory (HRT) of LRPM by Brognara
et al.\cite{brognara}, who also used the correct form of potential
in their calculations. Note, that both of these theories agree on
the tricritical density to be $\rho^*_{tri}\approx 0.38$,  which
is about 20\% lower than our simulation results. However, while
taking into account Bjerrum pairing of ions into neutral dipoles
\cite{bjerrum} in Debye-H\"uckel approach would supposedly produce
better results by shifting the phase coexistence to lower
temperatures and higher densities, the  predictions of the HRT
seem to be final.

The phase coexistence for different degrees of  anisotropy
$\alpha$ is shown in Fig.1. At any $\alpha$, the system shows only
order-disorder phase separations  and a tricritical point. Thus,
the effect of stretching the lattice amounts merely to altering
the tricritical point location, without changing qualitatively the
topology of the phase diagram. This goes in contrast with the
recent theory based on Debye-H\"{u}ckel calculations
\cite{kobelev02-2} which predicts that at the lattice parameters
ratio $b/a \ge 2.6$ a distinct gas-liquid phase coexistence may
reappear.  However, as was mentioned above, the theoretical
approach of Ref.\cite{kobelev02-2}  disregards  the Bjerrum
pairing \cite{bjerrum}, or, more generally, clustering of ions,
which is  crucial for the thermodynamics of the gas-liquid
transition in electrolytes. Typically,  taking into account
pairing somewhat diminishes the critical temperature but
significantly increases the critical density
\cite{levin96,levin02,kobelev02-1}.  Nevertheless, taking into
consideration  the ion clustering  is unlikely to change the
topology of the phase diagram of LRPM on the isotropic lattice.
However, since the strength of the electrostatic interactions
increases with anisotropy, this may lead to more profound tendency
of ions to bind into neutral dipoles/clusters (mostly, in
$xy$-planes), producing eventually a Kosterlitz-Thouless (KT)
transitions \cite{kosterlitz} line at infinite stretching.
Stronger pairing then could shift the gas-liquid coexistence curve
much further than in the isotropic case, again putting it
completely inside the order-disorder phase envelope. It should be
noted that simulations of the two-dimensional lattice Coulomb gas
\cite{teitel97} also reveal no normal first order gas-liquid
transition, but only an order-disorder phase separations
\cite{teitel97}.

Tricritical temperature and density as functions of the lattice
parameters ratio $\alpha$ are presented in Fig.2. As anisotropy
increases, both tricritical temperature and density increase.
Physical account of this behavior follows from the properties of
the lattice Coulomb potential. Indeed, in $z$-direction lattice
potential (\ref{fiofr}) behaves almost as $1/z$ (see Table 1).
However, in crystallographic directions different from $[001]$,
the anisotropic lattice potential is much stronger than $1/r$ and
decays slower. While the interactions in $z$-direction almost do
not change with $\alpha$, stronger anisotropy brings stronger
interactions inside the $xy$-layers.  This becomes especially
important at large densities, where the average distance between
ions is of the order of the lattice spacing. Therefore, the
average energy per particle increases with anisotropy. As a
consequence, higher temperatures are required for fluctuations to
achieve a critical level, and the tricritical temperature grows.
At the same time, more intensive interactions between ions lead to
more active clustering. The free charges density, and hence the
electrostatic part of the free energy diminishes significantly.
However, the entropic part of the energy remains the same.
Therefore, electrostatic interactions of two charged sublattices
become comparable with the entropy of charge separation at higher
overall densities, and the coexistence shifts to denser phases.

The trend for tricritical temperature and density to increase
with anisotropy is correctly reproduced by the Debye-H\"uckel
treatment of anisotropic lattices \cite{kobelev02-2}: see Fig.2.
However, the quantitative agreement between theoretical
predictions and simulations is not very good. While the theory
predicts only a few percent increase in density at large lattice
stretching, our simulations showed more significant growth. For
the tricritical temperature, on the contrary, simulations yield a
somewhat smaller slope.

At large stretching, increasing interactions inside  $xy$-planes
result in lesser relative importance of interactions between ions
with different $z$ coordinates. At some point, interactions with
ions in the same layer start dominating, and the $3D$ system
becomes quasi-$2D$. Since different layers become uncoupled, due
to its probabilistic nature phase separation in each of them takes
place at different values of the chemical potential, and one may
have configurations with gas phase in some of the layers and
high-density ordered phase in the others, which prevents reliable
sampling of the system. At $\alpha$=10, it turned to be impossible
to obtain a reliable phase separation in a periodic  system of
more than one layer in $z$-direction: at high densities the system
would not equilibrate after more than $10^{9}$ Monte Carlo steps.

To this end, for $\alpha=10$ we carried out simulations of the
system consisting of only {\it one} plane, that is, square box
with $L^*_x=L^*_y=32$ periodic in $x$ and $y$ directions, with the
potential (\ref{fiofr}). Using only one layer in simulations
yielded a highly asymmetric phase coexistence very similar to that
obtained by Teitel for Coulomb gas on two-dimensional square
lattice \cite{teitel97}, with the tricritical point located at
$T^*_{tri} = 0.28,\;\;\rho^*_{tri}=0.70$: see Fig.1.  But in two
dimensions, the coefficient in the Laplace equation
(\ref{poisson}) is not $4\pi$ but $2\pi$. Since all our
simulations were done with three-dimensional potential, the
temperature at large anisotropies must be divided by 2 to obtain
the correct $2D$ limit, while the density should remain intact.
This yields $T^*_{2D}=0.14, \; \rho^*_{2D}=0.70$ which exactly
coincides with the values obtained by Teitel.

It should be also noted that parts of the coexistence curve for
$\alpha=10$ at intermediate densities lay below the corresponding
curve for $\alpha=4$. However, for the latter case in contrast to
$\alpha=10$, the interactions between layers are still very
important and add significantly to the total free energy of the
system: simulations of periodic system of only one layer produce
much lower coexistence curve. Therefore, although at equal
densities the energy of ions interactions inside a single layer
for $\alpha=4$ is less than for $\alpha=10$, the total energy
density may be higher because of the interactions between layers,
and hence the temperature of phase separations would also be
higher. However, this is not essential at very low or high
densities where the interactions inside layers dominate anyway.
Taking into account the influence of layers on each other at
$\alpha=10$ probably would eliminate this intersection of the
coexistence curves. Nevertheless, this hardly would change our
estimate for the tricritical point location in the limit of
infinite stretching, which, as it has been mentioned, coincides
with the results by Teitel for two-dimensional square lattice.

To gain further insight on the behavior of  LRPM on anisotropic
lattices we also performed simulations with the continuous space
Coulomb potential. The corresponding phase diagrams for box size
$L^*=16$ and $\alpha\ge 2$ are presented in Fig.3. At any fixed
lattice stretching, the phase coexistence is very similar to the
case of the true lattice potential. No normal gas-liquid phase
transition is found, and the phase coexistence  is between a
disordered low-density phase and ordered high-density phase. The
tricritical temperature shows the same trend as for the correct
lattice potential, increasing with the inter-layer distance.
However, since for small $\alpha$ the interactions are weaker now,
the tricritical temperatures are lower.

The scaling of the temperature $T$ with $E_0=q^2/Db$ is maintained
in this paper to allow comparisons with earlier theoretical
results. For the correct lattice (or continuum) two-dimensional
potential, this is allowed as $b\to\infty$ because the potential
becomes length scale-independent at the two-dimensional limit.
However, for the $1/r$ potential, using a scaling with the
layer-layer energy results in a divergence of the reduced critical
temperature. While for the correct lattice potential the "natural"
scale of energy is set by layer-layer interactions whose
asymptotic behavior does not change with $\alpha$ (see Table 1),
for the $1/r$ potential this scale is given by the energy of the
ion-ion interactions at contact {\it inside} a layer
$E^\sharp_0=q^2/Da$. Note, however, that for actual simulations
this is not important, because all what matters is the Boltzmann
factor, which remains the same as long as the energy and the
temperature are measured in the same units.

Calculations with the temperature scaled by this energy
$T^\sharp=kTDa/q^2$ (where $D$ is the three-dimensional dielectric
constant, appropriate for the $1/r$ potential) are shown in Fig.
4. With this "natural" scaling, the tricritical point behavior as
a function of anisotropy is drastically different: the tricritical
temperature {\it decreases} with anisotropy, converging to
$T^\sharp = 0.11$. The reason for this is that now increasing
anisotropy is not accompanied by stronger interactions between
ions. On the contrary, while interactions inside the $xy-$planes
remain the same, due to a weaker coupling of the layers effective
energy per particle diminishes, and so does the tricritical
temperature. This decay is rather fast and the temperature
saturates already at the lattice parameters ratio $\alpha \simeq
2$.  Detailed studies of ions configurations showed that at
$\alpha \ge 2$ different $xy-$planes become totally independent of
each other, and the system behaves as a quasi-$2D$ liquid.
Similarly to the true lattice potential case, to obtain phase
coexistence for the limit of infinite stretching ($\alpha\ge 2$),
simulations of periodic system of only one plane were carried out
with box size $L=32$ and $L=48$.

The tricritical density gradually increases from
$\rho^*_{tri}\simeq 0.48$ for isotropic lattice to
$\rho^*_{tri}\simeq 0.55$ at strong anisotropy, following the
trend shown by the system with correct lattice interactions: see
Fig.3. This again reflects the relative importance of clustering
when one approaches the two-dimensional limit of infinite
stretching. However, since the potential is now weaker, the
limiting density is lower than in true lattice $2D$ systems.

\section{Conclusions}

In this study, grand canonical Monte Carlo simulations and
histogram reweighting techniques have been used to investigate
phase behavior of the restricted primitive model with correct
lattice Coulomb potential on simple cubic and tetragonal lattices.
Our results show that, at all degrees of anisotropy, only
order-disorder phase separations and a tricritical point exist.
For isotropic cubic lattice, the effect of using the correct
lattice potential shows only in the value of the tricritical
temperature.  As anisotropy increases, both  the tricritical
temperature and density  increase.  At large anisotropy  the
system undergoes a qualitative shift from a three-dimensional to
two-dimensional behavior. This shift is not accompanied by any
change in the phase diagram topology, and only the tricritical
parameters alter. To study in detail the transition from $3D$ to
$2D$ behavior, further investigations are needed. In particular,
its location may depend on the system size and boundary conditions
used in Ewald summation. For continous $1/r$ potential, the
tricritical temperature displays an opposite trend and decreases
with the anisotropy. Although exact values of the tricritical
temperature and density can be influenced by finite size effects,
an issue totally omitted in this study,  their qualitative
behavior seems physically plausible. It would be also  useful to
develop  another procedure for Ewald-like summation  for lattice
potential, which will not not rely on its asymptotic behavior.

The only existing theoretical treatment of Coulomb gas on
anisotropic  lattices based on the Debye-H\"uckel approach
captures correctly the trend of the tricritical temperature and
density  behavior with anisotropy \cite{kobelev02-2}.  However,
the absence of the anticipated  normal gas-liquid transition
suggests that with increasing anisotropy, stability of both
ordered and disordered dense phases increases in a similar way,
and the formation of an ordered structure remains
thermodynamically more favorable than possible gas-liquid phase
coexistence.  Taking explicitly  into account the ion clustering
will provide a better theoretical description of thermodynamics of
ionic systems on anisotropic lattices.

\section*{Acknowledgements}
Acknowledgment is made to the Donors of the American  Chemical
Society Petroleum Research Fund (Grants No. 37867-G6 to ABK and
38165-AC9 to AZP) for support of this research. ABK also
acknowledges the support of the Camille and Henry Dreyfus New
Faculty Awards Program (Grant No. NF-00-056). The authors would
like to thank Professor M.E. Fisher for critical comments and
useful suggestions.

\begin{appendix}
\section{ Lattice Coulomb Potential: Exact Representation and Fast Calculation}

Due to the discrete symmetry, the correct  lattice Coulomb
potential differs from continous $1/r$. The analytical expression
for this potential  follows from the lattice version of the
Poisson equation \cite{kobelev02-1},
\begin{equation} \label{poisson}
\Delta \varphi({\bf r}) = -\frac{4\pi}{D v_0}\delta({\bf r}),
\end{equation}
where the exact form of the lattice Laplacian  depends on the
geometry of the lattice, $D$ is the dielectric constant of the
media, and $v_0$ is the unit cell volume. For simple tetragonal
lattice with lattice spacings $a$ in $x$ and $y$ directions and
$b$ in $z$ direction  one has
\begin{equation} \label{laplac1}
    \Delta \varphi= \Delta_x \varphi+\Delta_y \varphi+\Delta_z \varphi,
\end{equation}
with
\begin{equation} \label{laplac2}
   \Delta_i \varphi(\mathbf{r}) = 1/a_i^2 \left[\varphi(\mathbf{r}-a_i\mathbf{e}_i)-2\varphi(\mathbf{r})+\varphi(\mathbf{r}+a_i\mathbf{e}_i)\right],
\end{equation}
where $i=x,y,z$; $a_x=a_y=a,  \ a_z=b$ and $\mathbf{e}_i$  are the
unit vectors along the corresponding lattice directions. After
introducing the anisotropy parameter $\alpha = b/a$, the solution
of the Poisson equation has the following form
\begin{equation} \label{phi}
  \varphi(\mathbf{r})=\frac{1} {4\pi^2 Db}\int_{-\pi}^{\pi}  \frac{e^{i \mathbf{k r}}d^3\mathbf{k}}{2+1/\alpha^2 - (\cos k_1 + \cos k_2 + 1/\alpha^2 \cos k_3)}.
\end{equation}

However, the potential in the form of a triple integral  of a
periodic function is not convenient for numerical calculations.
More practical expression for this integral has been  obtained by
Maradudin et al. \cite{maradudin}
\begin{equation} \label{bessel}
     \varphi(\mathbf{r}\equiv(l,m,n)) = \frac{2\pi}{Db}  \int_0^\infty \exp[-(2+1/\alpha^2)t] I_l(t)I_m(t)I_n(t/\alpha^2) dt,
\end{equation}
where $I_l$ denotes a modified Bessel function of the first kind.

To enhance calculations of the lattice potential for the  whole
lattice, an error-free propagating algorithm, similar to that
proposed by Friedberg and Martin \cite{friedberg} for isotropic
lattice, has been used. First, we compute lattice potentials in
planes $\langle x0z\rangle$ and $\langle x1z \rangle$ using
Eq.(\ref{bessel}). Now, for a given lattice cite ($k,l,m$), the
anisotropic lattice potential satisfies the gradient equations
\begin{eqnarray} \nonumber \label{gradient}
\frac{\varphi(l,m,n)-\varphi(l-1,m,n)}{l} &=& \frac{\varphi(l,m+1,n)-\varphi(l,m-1,n)}{m}\\
 &=&\alpha^2\frac{\varphi(l,m,n+1)-\varphi(l,m,n-1)}{n}
\end{eqnarray}
which follow from Eq.(\ref{phi}). At fixed $n$, potential  at any
point $(l,m,n)$ can be obtained recursively from two lines, $m=0$
and $m=1$. For $l=0$ it would be indeterminate, but this is not
important  since due to the symmetry
$\varphi(0,m,n)=\varphi(m,0,n)$. As Friedberg and Martin showed
\cite{friedberg}, this recursion does not lead to increase of the
initial possible rounding error at $m=0$ and $m=1$ lines as long
as $|l|\le|m|$. For $|l|>|m|$ the symmetry
$\varphi(l,m,n)=\varphi(m,l,n)$ can be employed.

Note that similar considerations can be done for  initial $\langle
x0z\rangle$ and $\langle x1z\rangle$ planes in attempt to generate
the potentials from two lines, using the second equation in
(\ref{gradient}). In these planes the potential is not symmetric,
and therefore two sectors, $|l|\le\alpha^2|n|$ and
$|l|\le\alpha^2|n|$ need to be calculated separately, starting
from the corresponding axes and then "stitched" together at the
$|l|=\alpha^2|n|$ line. However, since all potentials and
interactions are computed only once at the beginning of
simulations and then fast look-up algorithm is used, this gives us
a relatively small advantage in the overall computational time.

\end{appendix}

\newpage

\begin{figure}
\unitlength 1in
\begin{picture}(4,4)
\resizebox{4in}{4in}{\includegraphics{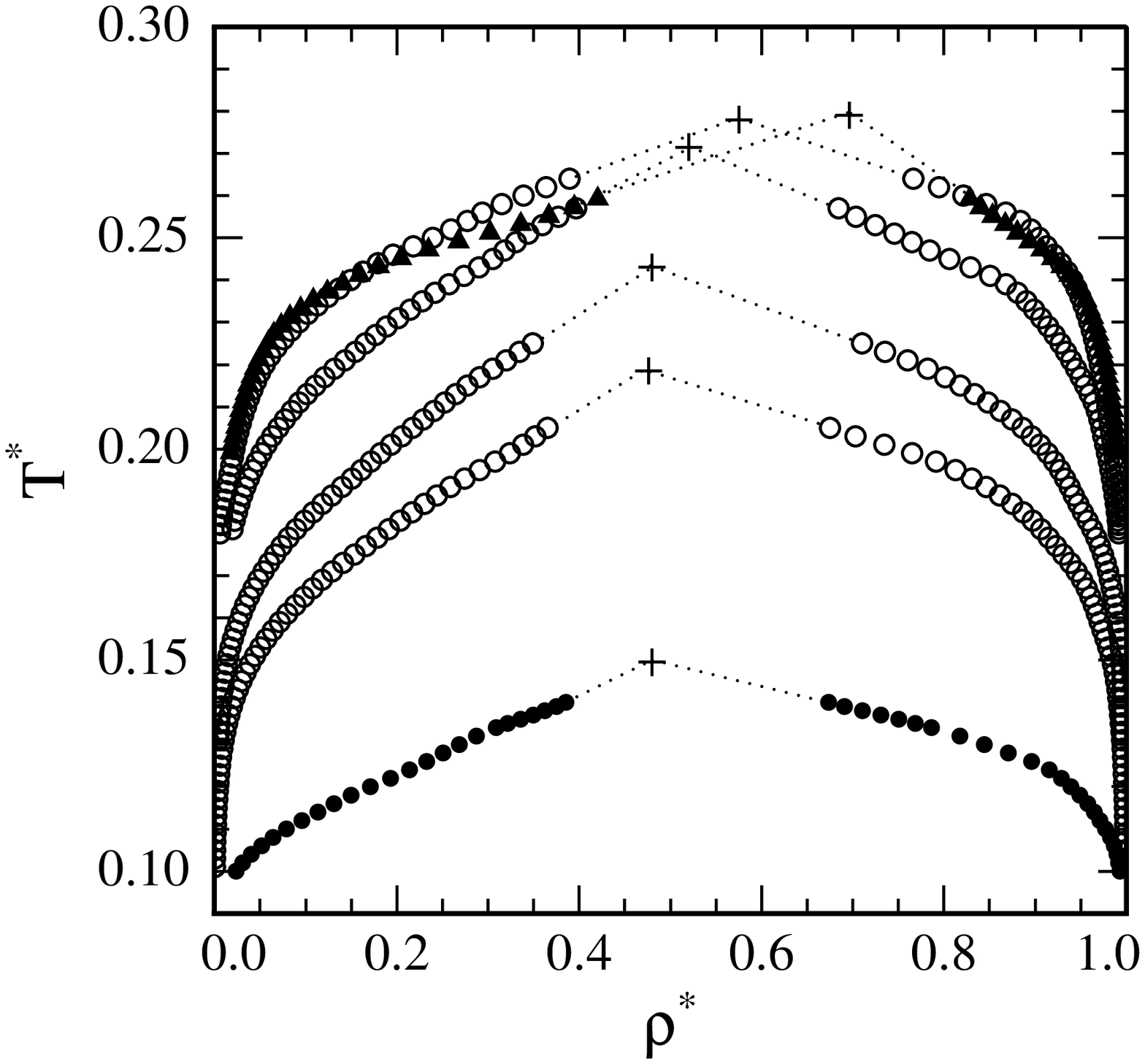}}
\end{picture}

\caption{Phase diagrams in LRPM with correct  lattice potential on
tetragonal lattices for different degrees of anisotropy (open
circles). Open circles from  bottom to top $\alpha=1,1.2,2,4$;
solid triangles correspond to $\alpha$=10. For reference, phase
coexistence in LRPM on cubic lattice with $1/r$ potential
\cite{panagiotopoulos99} is shown in solid circles (see also
Fig.3). }

\end{figure}

\begin{figure}
\unitlength 1in
\begin{picture}(4,4)
\resizebox{4in}{4in}{\includegraphics{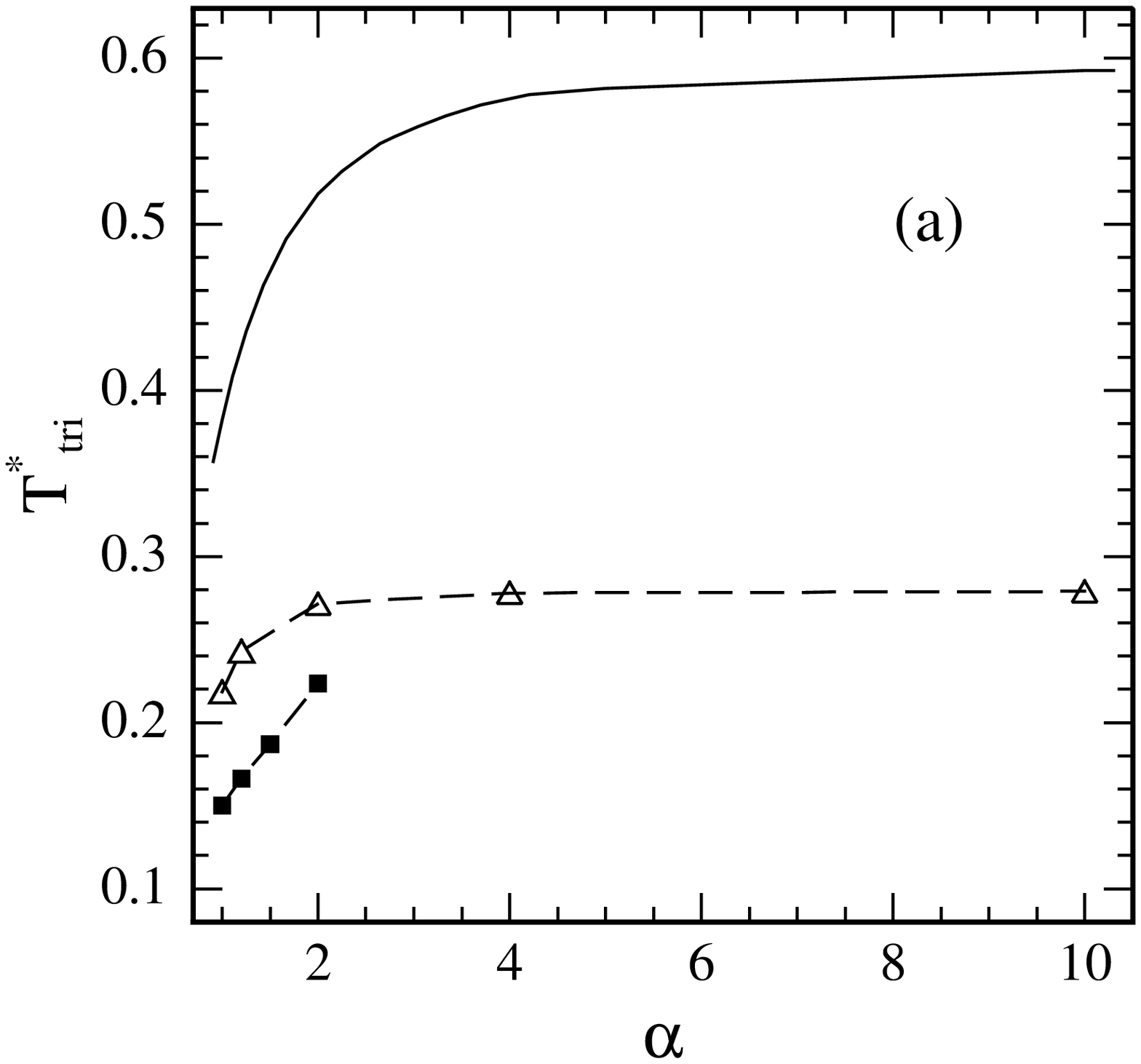}}
\end{picture}
\begin{picture}(4,4)
\resizebox{4in}{4in}{\includegraphics{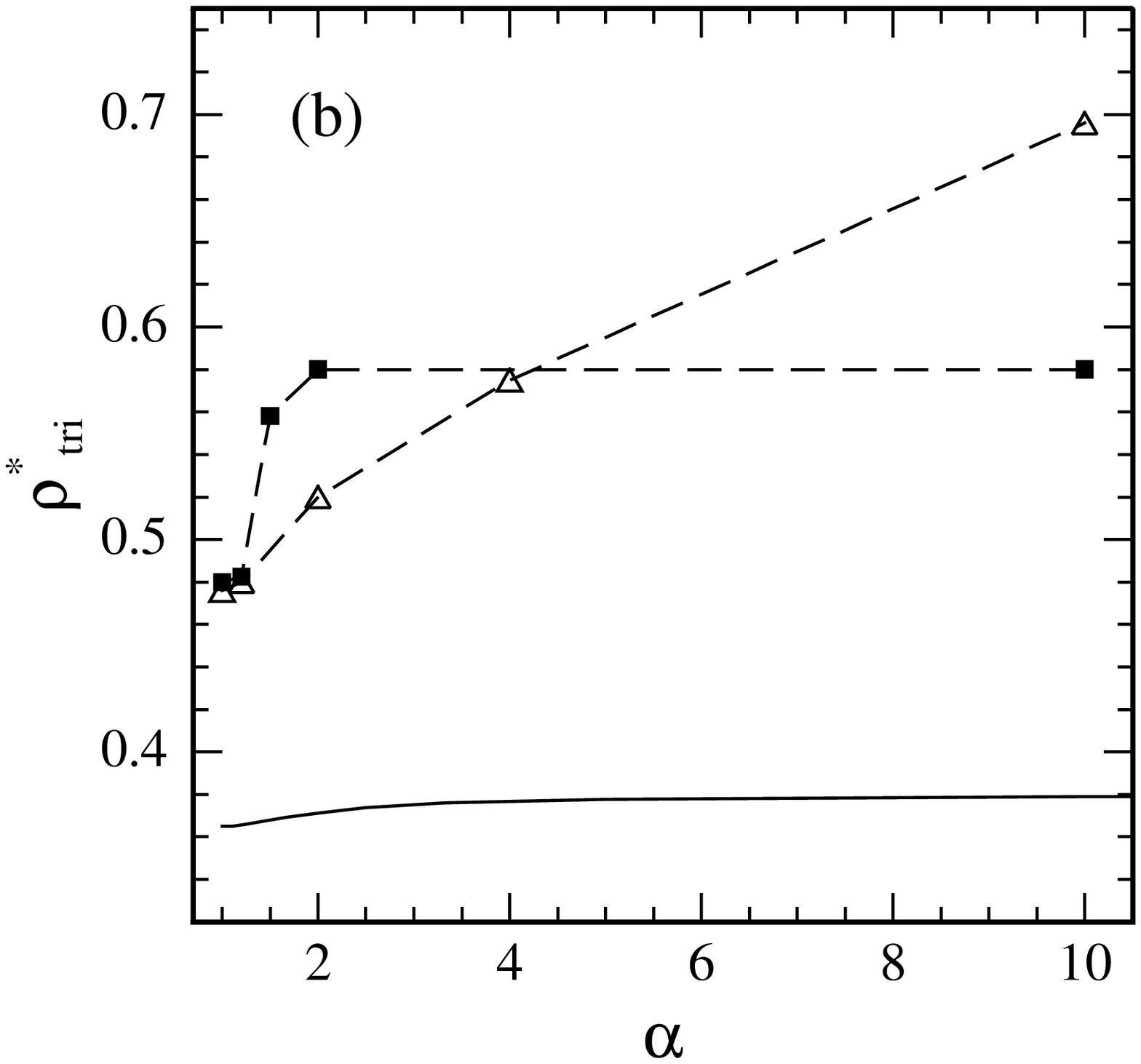}}
\end{picture}
\caption{Reduced tricritical parameters as a  function of the
lattice spacing ratio. (a) tricritical temperature, (b)
tricritical density. Open triangles correspond to the correct
lattice potential, squares are for the continuous space potential
$1/r$. Dashed lines are merely guides to the eye. The analytical
theory predictions are shown in solid lines \cite{kobelev02-2}.}

\end{figure}

\begin{figure}
 \unitlength 1in
\begin{picture}(4,4)
\resizebox{4in}{4in}{\includegraphics{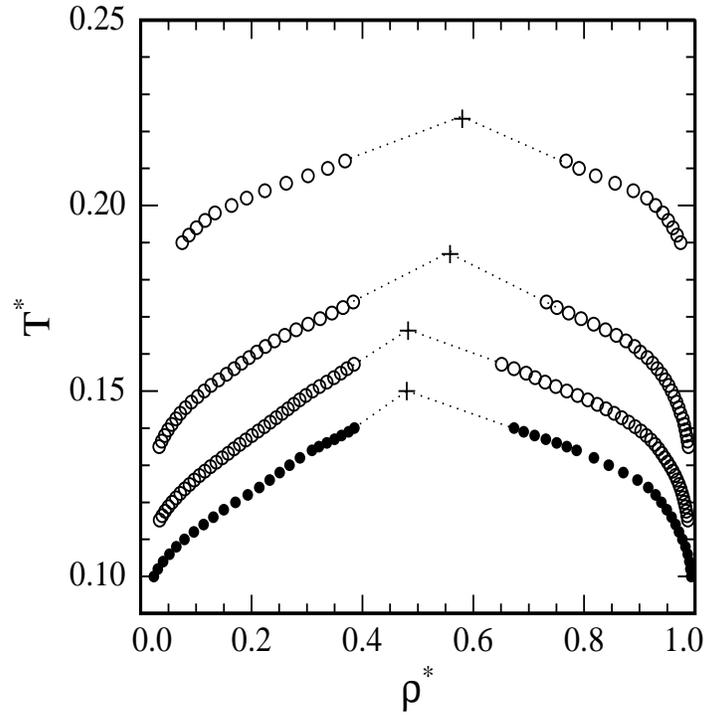}}
\end{picture}
\caption{Phase diagrams of LRPM with  continuous-space $1/r$
Coulomb potential. From bottom to top, $\alpha=1,1.2,1.5,2$.
Reduced temperature is defined by Eq.(2).}

\end{figure}

\begin{figure}
 \unitlength 1in
\begin{picture}(4,4)
\resizebox{4in}{4in}{\includegraphics{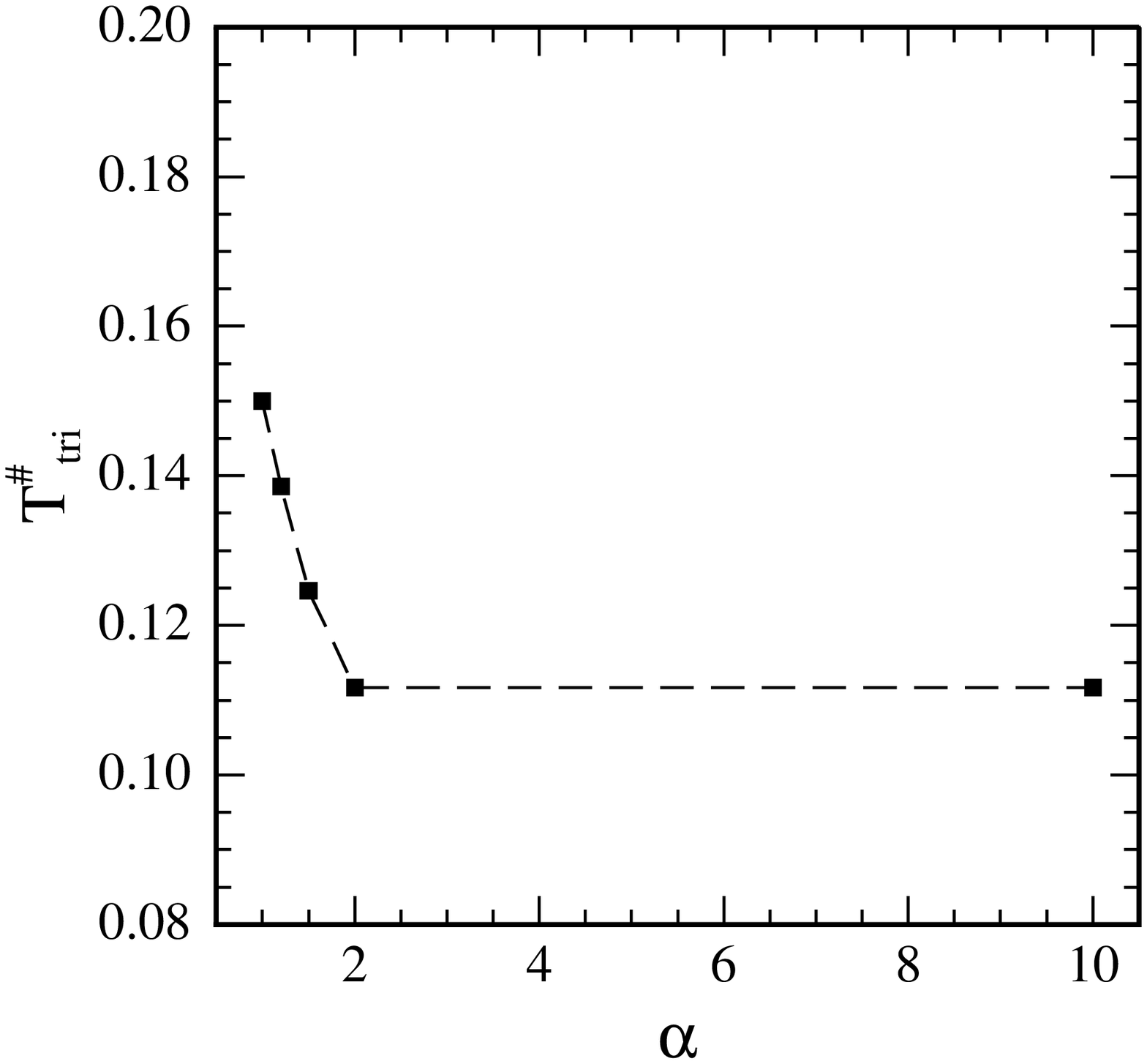}}
\end{picture}
\caption{ Reduced tricritical temperature $T^\sharp=kTDa/q^2$ for
the system with continuum $1/r$ Coulomb potential as a function of
the anisotropy.}

\end{figure}


\begin{references}

\bibitem{pitzer}  K.C. Pitzer, Acc. Chem. Res. \textbf{23}, 373 (1990).

\bibitem{japas} M.L. Japas and J.M.H. Levelt-Sengers, J. Phys. Chem. \textbf{94}, 5361 (1990).

\bibitem{weingartner01} H. Weing\"artner and W. Schr\"oer, Adv. Chem. Phys. \textbf{116}, 1 (2001).

\bibitem{levin02} Y. Levin, Rep. Progr. Phys. \textbf{65}, 1577 (2002).

\bibitem{levin96}  M.E. Fisher and Y. Levin, Phys. Rev. Lett. \textbf{71}, 3826 (1993); Y. Levin and M.E. Fisher, Physica A  \textbf{225}, 164 (1996).

\bibitem{stell95} G. Stell, J. Stat. Phys. \textbf{78}, 197 (1995).

\bibitem{yeh96} S. Yeh, Y. Zhou and G. Stell, J. Phys. Chem. \textbf{100}, 1415 (1996).

\bibitem{blum02} J. Jiang, L. Blum, O. Bernard, J. Prausnitz and  S.I. Sandler, J. Chem. Phys. \textbf{116} 7977 (2002).

\bibitem{kulinskii} V.L. Kulinskii and N.P. Malomuzh, Phys. Rev. E \textbf{67} 011501 (2003).

\bibitem{panagiotopoulos99} A.Z. Panagiotopoulos and S.K. Kumar, Phys. Rev. Lett \textbf{83}, 2981 (1999); A.Z. Panagiotopoulos,  J. Chem. Phys. \textbf{116}, 3007 (2002).

\bibitem{caillol02} J.-M. Caillol, D. Levesque and J.-J Weis, J. Chem. Phys. \textbf{116}, 10794 (2002).

\bibitem{lujiten02}  E. Luijten, M.E. Fisher and A.Z. Panagiotopoulos, Phys. Rev. Lett. \textbf{88}, 185701 (2002).

\bibitem{zuckerman01} D.M. Zuckerman, M.E. Fisher and S. Bekiranov, Phys. Rev. E \textbf{64}, 011206 (2001).

\bibitem{banerjee02} S. Banerjee and M.E. Fisher, (to be published).

\bibitem{artyomov03} M.N. Artyomov, V. Kobelev and A.B. Kolomeisky, J. Chem. Phys. {\bf 118}, 6394 (2003).

\bibitem{netz} R.R. Netz and H. Orland, Europhys.Lett. {\bf 45}, 726 (1999).

\bibitem{sabir} A.K. Sabir, L.B. Bhuiyan and C.W. Outhwaite, Mol.Phys. {\bf 93}, 405 (1998).
\
\bibitem{camp99} P.J. Camp and G.N. Patey, J.Chem.Phys. {\bf 111}, 9000 (1999).

\bibitem{panagiotopoulos02} A.Z. Panagiotopoulos and M.E. Fisher, Phys.Rev.Lett {\bf 88}, 045701 (2002).

\bibitem{yan02} Q. Yan and J.J. de Pablo, J. Chem. Phys. {\bf 116} 2967 (2002)

\bibitem{cheong03} D. W. Cheong and A. Z. Panagiotopoulos, J. Chem. Phys., submitted (2003)

\bibitem{dickman99} R. Dickman and G. Stell, in \textit{Simulation and Theory of Electrostatic Interactions in Solutions}, ed. L.R. Pratt, G.Hummer  (AIP, Woodbury, 1999).

\bibitem{kobelev02-1} V. Kobelev, A.B. Kolomeisky and M.E. Fisher, J. Chem. Phys. \textbf{116}, 7589 (2002).


\bibitem{brognara} A. Brognara, A. Parola, L. Reatto, Phys. Rev. E {\bf 65}, 066113 (2002).

\bibitem{ciach01} A. Ciach and G. Stell, J. Chem. Phys. {\bf 114}, 382 (2001); J. Chem. Phys \textbf{114}, 3617 (2001).

\bibitem{kobelev02-2} V. Kobelev and A.B. Kolomeisky, J. Chem. Phys. {\bf 117}, 8879 (2002).

\bibitem{diehl03} A.Diehl and A.Z.Panagiotopoulos, J. Chem. Phys. \textbf{118}, 4993 (2003).


\bibitem{frenkel} D.Frenkel and B. Smit, {\it Understanding Molecular Simulations} (Academic, New York, 1996).

\bibitem{deLeeuw} S.W. deLeeuw, J.W.Perram and E.R. Smith, Proc. Roy. Soc. Lon. A, {\bf 373} 27 (1980).

\bibitem{friedberg} R. Friedberg and O. Martin, J. Phys. A: Math. Gen. {\bf 20}, 5095 (1987)

\bibitem{orkoulas94} G. Orkoulas and A.Z. Panagiotopoulas, J. Chem. Phys. {\bf 101}, 1452 (1994).

\bibitem{ferrenberg} A.M. Ferrenberg and R.H.Swendsen, Phys. Rev. Lett. {\bf 61}, 2635 (1988).

\bibitem{bjerrum} N. Bjerrum, K. Dan Vidensk. Selsk. Mat. Fys. Medd. \textbf{7}, 1 (1926).

\bibitem{kosterlitz} J.M. Kosterlitz, D. Thouless, J. Phys. C {\bf 6}, 1181 (1973); J.M. Kosterlitz,  J. Phys. C {\bf 7}, 1046 (1974)

\bibitem{teitel97} P. Gupta and S. Teitel, Phys.Rev.B {\bf 55}, 2756 (1997)


\bibitem{maradudin} A.A. Maradudin, E.W. Montroll, G.H. Weiss, R. Herman, and H.W. Milnes, {\it Green's Functions for monoatomic Simple Cubic Lattices} (Academie Royale de Belgique, Brussels, 1960); see also R.T.Delves and G.S. Joyce, Ann. Phys {\bf 291}, 71 (2001).

\end{references}
\end{document}